\documentclass[twocolumn,showpacs,preprintnumbers,amsmath,amssymb]{revtex4}

\usepackage{graphicx}

\begin{document}

\title{A spatial model of autocatalytic reactions}

\author{Pietro de Anna}
\affiliation{G\'eosciences Rennes, UMR 6118, CNRS, Universit\'e de Rennes 1,
Rennes, France}

\author{Francesca Di Patti}
\affiliation{Dipartimento di Fisica ``Galileo Galilei'', Universit\`{a} degli
Studi di Padova, via F. Marzolo 8, 35131 Padova, Italy }

\author{Duccio Fanelli}
\affiliation{Dipartimento di Energetica, University of Florence and INFN,
Via S. Marta 3, 50139 Florence, Italy}

\author{Alan J. McKane}
\affiliation{
Theoretical Physics, School of Physics and Astronomy, University of Manchester,
Manchester M13 9PL, United Kingdom}

\author{Thierry Dauxois}
\affiliation{Universit\'e de Lyon, Laboratoire de Physique de l'\'Ecole
Normale Sup\'erieure de Lyon, CNRS, France}

\begin{abstract}
Biological cells with all of their surface structure and complex interior
stripped away are essentially vesicles --- membranes composed of lipid
bilayers which form closed sacs. Vesicles are thought to be relevant as 
models of primitive protocells, and they could have provided the ideal 
environment for pre-biotic reactions to occur. In this paper, we investigate 
the stochastic dynamics of a set of autocatalytic reactions, within a 
spatially bounded domain, so as to mimic a primordial cell. The 
discreteness of the constituents of the autocatalytic reactions gives rise 
to large sustained oscillations, even when the number of constituents is
quite large. These oscillations are spatio-temporal in nature, unlike those 
found in previous studies, which consisted only of temporal oscillations. 
We speculate that these oscillations may have a role in seeding membrane 
instabilities which lead to vesicle division. In this way synchronization 
could be achieved between protocell growth and the reproduction rate of 
the constituents (the protogenetic material) in simple protocells.
\end{abstract}

\pacs{02.50.Ey, 05.40.-a, 87.16.dj}

\maketitle

\section{Introduction}

The cell is a structural and functional unit, the building block of any
living system. Cells consist of a membrane, made of a lipid bilayer, which
encloses and protects the contents of the cell, including genetic material.
The membrane is semi-permeable: nutrients can diffuse in and serve as energy
to support the functioning of the machinery~\cite{alb02}. Cells undergo
replication (cell division): this is a process by which a cell, hereafter
called the parent cell, divides into two or more cells, called the
daughters. The daughter cell contains in principle an exact replica of the
parent's inner constituents, this property being ultimately a prerequisite
for stable living organisms to exist. Such a process clearly relies on the
synchronization between the duplication rate of the constituents and the growth
of the container. In modern cells this condition is of paramount importance and
is efficiently realized via dedicated control mechanisms, expressed as
pathways of nested molecular checkpoints~\cite{alb02}. This complex and
delicate machinery has evolved; presumably the first minimalistic cells
(so-called protocells~\cite{dea86}-\cite{ras08}) had a far more straightforward
and less elaborate way of dividing. So focusing on the primordial cell units
postulated to be present at the origin of life on Earth, can we conceive of
a simple, though efficient, mechanism which could govern the division process?
A possible answer to this question will emerge as a result of the calculations
carried out in this paper.

One of the most persuasive scenarios concerning the origin of life on Earth
identifies vesicles as protocells~\cite{lui06}. These are tiny closed sacs 
in which the outer membrane takes the form of a lipid bilayer, and so 
are good candidates for a minimal cell. Despite the dramatic reduction in
complexity as compared to modern cells, vesicles still display many
fascinating properties, as revealed in laboratory
experiments~\cite{sei97,lui06}. They are semi-permeable and allow for different
types of chemicals to enter the enclosed volume, and so sustain any reaction
cycles that may be taking place. In addition, vesicles can grow due to
inclusion of lipid constituents into their surface, progressively adjust their
shape, and eventually divide to produce daughter vesicles. Vesicles which are
initially spherical can pass through a number of intermediate shapes before
they divide, for instance a vesicle may first change into an ellipsoid, then
into a dumbbell shape and finally into two attached spheres, at which point 
it will divide in two~\cite{sei97}.

However it must be said there is in reality very little theoretical evidence
that the shape of the vesicle always follows this particular sequence, and
even less experimental evidence. It may be more appropriate to talk about
an ensemble of vesicles and typical pathways to the state where division takes
place. Similarly there may only be a mean time to division, although it should
be noted that there would then be a selection process which would favor the
types of vesicles (if they could be distinguished) which would undergo the
division proceeding at the fastest rate.

When modeling protocells, one needs to relate the mechanism of growth
and division to the actual microscopic dynamics of the internal
constituents. While vesicles can possibly define the scaffold of
prototypical cell models, what can one say about the internal constituents?
It is customarily believed that autocatalytic reactions~\cite{gra85} might
have had a role in producing complex molecules required for the origin of
life~\cite{dys85}-\cite{wac90}. A chemical reaction is called autocatalytic
if one of the reaction products is itself a catalyst for the chemical
reaction. Clearly the reaction will speed up as more catalyst is produced. If
there are several catalytic reactions, rather than just one --- an
autocatalytic set~\cite{jai98} --- then more complex behavior is possible,
with some reactions producing catalysts for other reactions. This suggests
that the interior of the protocell might have been occupied by interacting
families of replicators, organized in autocatalytic cycles.

Autocatalytic reactions have also been invoked in the context of
studies on the origin of life as a possible solution of the famous
Eigen paradox~\cite{eig71}. This is a puzzle, since it limits the
size of self-replicating molecules to perhaps a few hundred base
pairs. At odds with this conclusion, almost all life on Earth
requires much longer molecules to encode their genetic information.
This problem is dealt with in living cells by the presence of
enzymes which repair mutations, allowing the encoding molecules to
reach sizes on the order of millions of base pairs~\cite{lod99}. In
primordial organisms, autocatalytic cycles might have provided the
required degree of microscopic cooperation to prevent Eigen's
evolutionary drive to self-destruction to occur.

In this paper we will investigate the properties of autocatalytic
reactions within a bounded region of space, which we will identify
with the vesicle, the whole structure being a reference model for a
protocell. The autocatalytic reactions will be taken to have the form 
proposed by Togashi and Kaneko~\cite{tog01,tog03}. In their work, Togashi
and Kaneko emphasized the role played by the noise intrinsic to the
system of elementary constituents. This model was recently revisited
by Dauxois et al.~\cite{dau09}, who used an approach based on
expanding the master equation in a system-size
expansion~\cite{van07}, to make analytic progress in the description
of the process. This approach has recently been applied to a number
of processes in biological systems to show how large oscillations
can emerge, sustained by the stochastic component of the
dynamics~\cite{mck05}-\cite{alo07}. The analysis has also been
extended to a spatial model~\cite{lug08}, and our calculations will
mirror those in this paper.

Therefore here we will ask what happens once space (i.e. microscopic particle 
diffusion) is incorporated into the model. Are the oscillations robust or, 
conversely, do they get washed out through coarse-grained averaging? We shall
demonstrate that spatio-temporal patterns do emerge and influence the mass
transport inside the cell. We will also speculate that the division of the
protocell requires an inherent degree of synchronization which may be
triggered by collective, spatially ordered fluctuations in the concentration.
Building on this scenario, one can imagine that localized peaks in the
concentration might develop at a given stage of the vesicle evolution. Denser
patches could then drive an instability which could potentially lead to the
distortion of the membrane and so to division. 

\section{The model}
\label{model}

The model we will use is a spatial version of the autocatalytic model discussed
in~\cite{dau09}. The idea is to introduce a spatial coarse-graining and divide
the vesicle into small micro-cells, within which autocatalytic reactions
occur. The cells adjoining the membrane which forms the limit of the vesicle
have a special status, since the membrane allows chemicals to diffuse in from
the environment and out into the environment. In this paper we will focus only
on these micro-cells --- those that are adjacent to the boundary --- and
lump all the interior micro-cells together into an inner region. We do not
give the environment or this inner region any spatial structure; they simply
act as a particle reservoir for the chemicals in the micro-cells adjacent to
the membrane.

\begin{figure}[b]
\vspace{2truecm}
\includegraphics[width=8cm]{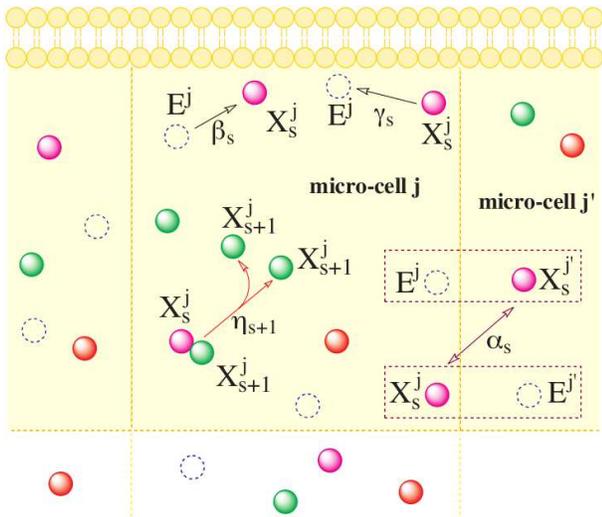}
\caption{(Color online) The volume of the cells adjacent to the boundary 
is imagined to be partitioned into $\Omega$ micro-cells (see also 
Figure~\ref{figure2}). Within micro-cell $j$ the molecular species interact 
according to the autocatalytic reactions specified by Eqs.~(\ref{eq:chim1}). 
In addition, the molecules can migrate from micro-cell $j$ to its nearest 
neighbors, e.g. micro-cell $j'$, as depicted in the cartoon. A molecule of 
type $X_s^j$ (full circle) takes over a vacancy (dashed empty circle) of 
micro-cell $E^{j'}$, and so transforms into $X_s^{j'}$, leaving behind a 
vacancy $E^j$. Finally, the chemical can also diffuse in from the environment, 
a reaction that in turn implies changing $E^j$ into $X_s^{j}$. The opposite 
holds for molecules that diffuse out into the environment.}
\label{figure1}
\end{figure}

In each micro-cell autocatalytic reactions as specified in~\cite{dau09} occur,
see Fig.~\ref{figure1}. More specifically we consider $k$ chemical species,
here labeled $X^j_s$, with the index $s=1,\ldots,k$ labeling the species and
$j=1,...,\Omega$, the $\Omega$ micro-cells where the reactions occur. The
autocatalytic reactions take the form~\cite{dau09}
\begin{equation}
\label{eq:chim1}
X_{s}^j + X_{s+1}^j \stackrel{\eta_{s+1}}{\longrightarrow} 2 X_{s+1}^j.
\end{equation}
The reactions are taken to be cyclic, so that $X_{k+1}^j= X_{1}^j$.

\begin{figure}[b]
\vspace{2truecm}
\includegraphics[width=8cm]{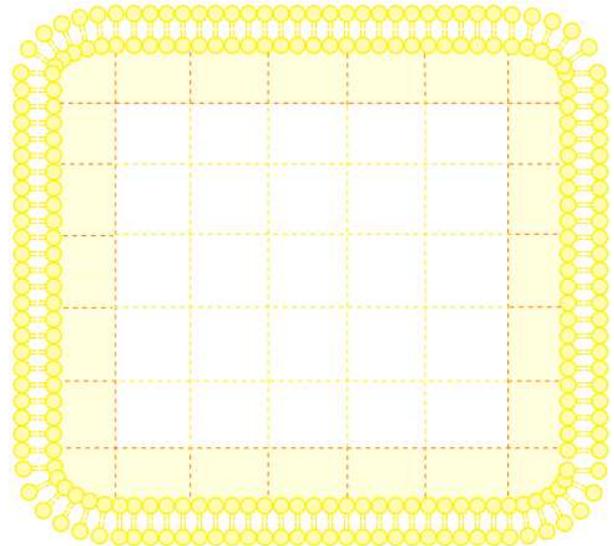}
\caption{(Color online) In the spatial autocatalytic model considered here 
the vesicle is imagined to be divided into small micro-cells. We are 
specifically interested in the micro-cells adjoining the membrane, shown
in darker outline in the figure. These latter link up together and constitute 
a sort of inner shell, immediately adjacent to the vesicle wall. Within each 
micro-cell the chemicals interact as shown in Figure \ref{figure1}.}
\label{figure2}
\end{figure}

The spatial element of the model is introduced through migration of chemical
species between neighboring cells. The boundary cells will form a periodic
structure in two dimensions, so that a Fourier-based approach can be used in
the analysis described below. The geometry is schematically depicted in
Fig.~\ref{figure2}, in a two-dimensional setting, so that the micro-cells
form a one-dimensional periodic structure. It should be emphasized that
although the scheme is illustrated in Fig.~\ref{figure2} with reference to
a two-dimensional vesicle for simplicity, the setting and analysis apply in
any spatial dimension including the relevant three-dimensional case. If the
vesicle is $d+1$-dimensional, clearly the micro-cells will form a
$d$-dimensional periodic structure.

The migration between adjacent cells is encapsulated in the following relations
\begin{equation}
\label{diff}
X_s^j + E^{j'} \stackrel{\alpha_s}{\longrightarrow} X_s^{j'} + E^j\,, \\
\end{equation}
\begin{equation}
\label{diff1}
E^j + X_s^{j'} \stackrel{\alpha_s}{\longrightarrow} E^{j'} + X_s^j\,,
\end{equation}
where $j$ and $j'$ label the adjacent cells and $E^i$ represents the number
of vacancies in cell $i$. We will assume that the capacity of each cell is
$N$, so that sum of the number of molecules of each species plus the number
of vacancies equals $N$ for every cell.

Finally, cell $j$ may lose a molecule $X^j_s$ to the environment or inner
region leaving a vacancy $E^j$ or a gain of a molecule $X^j_s$ from the
environment or inner region, i.e.
\begin{equation}
X_{s}^j \stackrel{\gamma_{s}}{\longrightarrow} E^j; \qquad
E^j \stackrel{\beta_{s}}{\longrightarrow} X_{s}^{j}.
\end{equation}
There is no need to distinguish between the environment and inner region; the
rates $\gamma_s$ and $\beta_s$ can simply be regarded as the combined rates
for both processes. In the rest of the paper we will simply refer to both
these regions as ``the environment''.


In the following we will formulate the model in terms of a chemical master
equation and find the mean-field solution as well as determining stochastic
corrections to this which occurs when $N$ is finite. We will also simulate
the stochastic dynamics and compare the results with the analytic formulas
we obtain.

To describe the model as a chemical master equation, we denote the number
of molecules of chemical species $s$ in cell $j$ by $n_s^j$, and so the state
of the system can be characterized by the vector
$\textbf{n}=(\textbf{n}^1,\textbf{n}^2,...,\textbf{n}^{\Omega})$ where
$\textbf{n}^j=(n_1^j,n_2^j,...,n_k^j)$. The transition rate from one state
$\textbf{n}'$, to another $\textbf{n}$, is denoted by
$T(\textbf{n}|\textbf{n}')$ --- with the initial state being on the right.
For example, the transitions stemming from the autocatalytic cycles are
\begin{equation}
T(n_{s}^j-1,n_{s+1}^j+1|n_{s}^j,n_{s+1}^j) =
\frac{\eta_{s+1}}{\Omega}\frac{n_{s}^j}{N}\frac{n_{s+1}^j}{N},
\label{local_TRs}
\end{equation}
where within the brackets we have chosen to indicate only the dependence
on those species which are involved in the reaction. The transition
rates associated with the migration between adjacent micro-cells
take the form
\begin{eqnarray}
T(n_{s}^j-1,n_{s}^{j'}+1|n_{s}^j,n_{k}^{j'}) &=&
\frac{\alpha_s}{z\Omega}\frac{n_{s}^j}{N}\Big(1-\sum_{m=1}^k
\frac{n^{j'}_m}{N}\Big), \nonumber \\
T(n_{s}^j+1,n_{s}^{j'}-1|n_{s}^j,n_{s}^{j'}) &=&
\frac{\alpha_s}{z\Omega}\frac{n_{s}^{j'}}{N}\Big(1-\sum_{m=1}^k
\frac{n^{j}_m}{N}\Big),
\label{migration_TRs}
\end{eqnarray}
where $z$ is the number of nearest neighbors that each micro-cell has. Finally,
for the interaction with the environment, the transition rates are
\begin{displaymath}
 T(n_{s}^j-1|n_{s}^j) = \frac{\gamma_{s}}{\Omega}\frac{n_{s}^j}{N},
\end{displaymath}
\begin{equation}
 T(n_{s}^j+1|n_{s}^j) = \frac{\beta_{s}}{\Omega}
\Big(1-\sum_{m=1}^{k}\frac{n^j_{m}}{N}\Big).
\label{environ_TRs}
\end{equation}
In Eqs.~(\ref{migration_TRs}) and (\ref{environ_TRs}), explicit use has been 
made of the condition
\begin{equation}
\sum^{k}_{s=1} \frac{n^j_s}{N} + \frac{n^j_E}{N} = 1,
\label{cons_num}
\end{equation}
to eliminate $n^j_E$, the number of vacancies in cell $j$.

The system is intrinsically stochastic and may be described by the probability
density function, $P(\textbf{n}, t)$, which gives the probability of finding
the system in state $\textbf{n}$ at time $t$. The equation which governs the
dynamical evolution of $P(\textbf{n},t)$ is the master equation~\cite{van07},
which for the system under consideration here takes the form
\begin{eqnarray}
\frac{dP(\textbf{n},t)}{dt} &=& \sum^{\Omega}_{j=1} {\cal T}^{j}_{\rm loc}
P(\textbf{n},t) + \sum^{\Omega}_{j=1} \sum_{j' \in j} {\cal T}^{jj'}_{\rm mig}
P(\textbf{n},t) \nonumber \\
&+& \sum^{\Omega}_{j=1} {\cal T}^{j}_{\rm env} P(\textbf{n},t),
\label{master}
\end{eqnarray}
where the three terms on the right-hand side refer to the local terms for the
chemical reactions, the migration of chemical species between the micro-cells,
and the interaction with the environment, respectively. The notation $j' \in j$
means that the cell $j'$ is a nearest-neighbor of the cell $j$. The three
terms in the master equation can be expressed in a concise, but transparent,
form by introducing the step operator~\cite{van07} ${\cal E}^{\pm 1}_{s,j}$
defined by
\begin{equation}
{\cal E}^{\pm 1}_{s,j} f(\{ n^i_m \} ) = f(\ldots, n^j_{s} \pm 1,\ldots),
\label{steps_ops}
\end{equation}
where $f$ is an arbitrary function. The explicit forms for these three terms
are
\begin{eqnarray}
{\cal T}^{j}_{\rm loc}  & = & \sum^{k}_{s=1} \left( {\cal E}_{s,j}
{\cal E}^{-1}_{s+1,j} - 1 \right) T(n^j_s - 1, n^j_{s+1} + 1 | n^j_s
, n^j_{s+1}) \nonumber \\  && \label{local} \\
{\cal T}^{jj'}_{\rm mig} & = & \sum^{k}_{s=1} \left( {\cal E}_{s,j}
{\cal E}^{-1}_{s,j'} - 1 \right)
T(n^j_s - 1, n^{j'}_s + 1 | n^j_s , n^{j'}_s) \nonumber \\
& & + \sum^{k}_{s=1} \left( {\cal E}_{s,j'} {\cal E}^{-1}_{s,j} - 1
\right) T(n^{j'}_s - 1, n^{j}_s + 1 | n^{j'}_s , n^{j}_s)
 \nonumber\\ && \nonumber \\\label{migration} \\
{\cal T}^{j}_{\rm env} &=& \sum^{k}_{s=1} \left[ \left( {\cal
E}_{s,j}
- 1 \right) T(n^j_s - 1 | n^j_s) \right. \nonumber \\
& & +\left. \left( {\cal E}^{-1}_{s,j} - 1 \right) T(n^j_s + 1 |
n^j_s) \right], \label{environ}
\end{eqnarray}
where it is understood that the operator ${\cal E}^{\pm 1}_{s,j}$ also acts
on $P(\textbf{n}, t)$ when these expressions are substituted into
Eq.~(\ref{master}). In Eq.~(\ref{local}) the cyclic nature of the reactions
means that $n^j_{k+1}$ should be identified as $n^j_1$ and
${\cal E}^{\pm 1}_{k+1,j}$ should be identified as ${\cal E}^{\pm 1}_{1,j}$.
The explicit expressions for the transition rates are given by
Eqs.~(\ref{local_TRs})-(\ref{environ_TRs}). These, together with
Eqs.~(\ref{master})-(\ref{environ}) define the model.

The above description is exact; no approximations have yet been
made. At this stage we could also resort to direct numerical
simulations of the chemical reaction system by use of the Gillespie
algorithm~\cite{gil76,gil77}. This method produces realizations of
the stochastic dynamics which are formally equivalent to those found
from the master equation (\ref{master}). Averaging over many
realizations enables us to calculate quantities of interest. We will
discuss the results of performing such simulations in Section
\ref{secondorder}, but a very accurate approximation scheme exists
which can be used to investigate models of this type analytically.
This is the van Kampen system-size expansion~\cite{van07}. It is
effectively an expansion in powers of $N^{-1/2}$, which to leading
order ($N \to \infty$) gives the deterministic equations describing
the system, and which at next-to-leading order gives finite $N$
corrections to these. These latter corrections take the form of
\textit{linear} stochastic differential equations which can then be
analyzed straightforwardly, especially in the case when the
deterministic system has approached a stable fixed point. The method
is based on substituting the ansatz
\begin{equation}
\frac{n_s^j}{N} = \phi_s^j + \frac{1}{\sqrt{N}}\xi_{s}^{j},
\label{ansatz}
\end{equation}
into the master equation (\ref{master}). Here $\phi_s^j(t)$ is the solution to
the deterministic equation, and $\xi_s^j(t)$ is a stochastic term which is the
difference between the actual value $n_s^j/N$ and $\phi_s^j$ at time $t$.

We develop this approximation in the next two sections. In
Section \ref{firstorder} we carry out the analysis to leading order, finding
the deterministic equations and the relevant fixed point. In
Section \ref{secondorder} we carry through the calculation to next-to-leading
order, investigating the linear stochastic differential equations by taking
their Fourier transforms. The derivations of these equations is lengthy,
though straightforward, and the details of the expansion are provided in
Appendices \ref{VK} and \ref{MandB}.

\section{Leading order: the deterministic equations}
\label{firstorder}
In the limit where the number of molecules (including vacancies) in each
micro-cell, $N$, goes to infinity, the system becomes deterministic and
is governed by a set of ordinary differential equations. These are found by
substituting the ansatz (\ref{ansatz}) into the master equation (\ref{master})
and letting $N \to \infty$, after the introduction of a rescaled time
$\tau={t}/{(N\Omega)}$. The calculation is described in Appendix \ref{VK},
but the same equation can also be found by multiplying Eq.~(\ref{master}) by
$n^{i}_{r}$ and summing over all states $\textbf{n}$. Either way one obtains
the following equation for species $s$ in cell $j$
\begin{eqnarray}
\frac{d \phi_s^j}{d\tau}& = &\eta_s\phi_{s-1}^j \phi_s^j -
\eta_{s+1}\phi_s^j \phi_{s+1}^j \nonumber \\
&+& \alpha_s\Big(\Delta\phi_s^j (1- \sum_{m=1}^{k} \phi_m^j) +
\phi_s^j \sum_{m=1}^{k} \Delta \phi_m^j \Big) \nonumber \\
&+& \beta_{s}(1-\sum_{m=1}^{k} \phi_m^j) - \gamma_{s}\phi_{s}^{j},
\label{eq:first_order}
\end{eqnarray}
where $\Delta$ is the discrete Laplacian operator
$\Delta f_s^j = (2/z) \sum_{j' \in j} (f_s^{j'} - f_s^j)$. In the limit where
the size of the micro-cells tends to zero, these equations become partial
differential equations, with $\Delta$ becoming the familiar Laplacian operator.
In this respect, Eq.~(\ref{eq:first_order}) generalizes the results
of \cite{dau09} to the case of a spatially-extended system. When turning off
the migration mechanism between neighboring micro-cells, i.e. imposing
$\alpha_s=0$ for any species $s$, the spatial aspects drop out and one
formally recovers the ordinary differential equations given in \cite{dau09}.

To proceed with the analysis, and to make contact with the investigation 
carried out in \cite{tog01, dau09}, we shall now assign the same chemical 
parameters to all the species. The migration rate is the only exception to 
this, and may have a different value for each species. We will see later that 
this will be necessary in order to find spatio-temporal oscillations but also, 
as we will see shortly, a straightforward analysis is still possible if we 
maintain in $\alpha_s$, and none of the other parameters, an explicit 
reference to the index $s$. We will be concerned with finding the homogeneous 
solution of Eq.~(\ref{eq:first_order}), that is, the solution with no spatial 
variation. The homogeneous solution is found to be an attractor of the 
deterministic dynamics, even when the system is initially prepared in a 
non-homogeneous configuration. This observation follows from numerical 
simulations, but can in principle be made quantitative by investigating the 
stability of the homogeneous fixed point. This means that no gradient in 
concentration is allowed between neighboring micro-cells, once the asymptotic 
regime is attained. So, when searching for fixed points of the dynamics, one
can set the terms involving the Laplacian in Eq.~(\ref{eq:first_order}) to 
zero. Since the only dependence on $s$, appearing in $\alpha_s$, multiplies 
these terms, there is also no dependence remaining on the species type, $s$, 
and so the fixed points are both independent of $j$ and of $s$. Under these
conditions a unique fixed point for the concentration, $\phi^*$, is easily 
found to be
\begin{equation}
\phi^* = \frac{\beta}{k \beta + \gamma},
\label{stationary}
\end{equation}
for any $s=1,...,k$ and $j=1,...,\Omega$. The result (\ref{stationary}) is
identical to that obtained in \cite{dau09}, when dealing with the non-spatial
homologous model.

In \cite{dau09}, fluctuations for finite $N$ were shown to induce
regular temporal oscillations in the species populations, so
significantly altering the predicted deterministic dynamics. What is
going to happen in the present spatial context? In Section \ref{secondorder} 
we shall investigate this, the central point of the paper, by focusing 
on the next-to-leading order corrections in the van Kampen expansion.

\section{Next-to-leading order: the stochastic corrections}
\label{secondorder}
Equating the terms of next-to-leading order in the master equation, after
rescaling the time, leads to the Fokker-Planck equation (\ref{eq:FP_sp})
which governs the probability density function of the fluctuations.
This Fokker-Planck equation is formally equivalent to the following Langevin
equation~\cite{gar04,ris89}
\begin{equation}
\frac{d\xi^j_s}{d\tau}= \sum_{j',r} M_{sr}^{jj'}\xi^{j'}_{r}
+ \lambda^j_s(\tau),
\label{eq:langevin}
\end{equation}
where
\begin{equation}
\langle \lambda_s^j(\tau)\lambda_r^{j'}(\tau') \rangle=
\mathcal{B}_{sr}^{jj'}\delta(\tau - \tau'). \label{eq:lambda}
\end{equation}
The noise term, $\lambda^j_s(\tau)$, in Eq.~(\ref{eq:langevin}) is Gaussian 
with zero mean and with a correlator given by Eq.~(\ref{eq:lambda}), from 
which it can be seen to be white. The form of the two  matrices $M$ and 
$\mathcal{B}$ are discussed in Appendix \ref{MandB}. They depend on the 
solution of the deterministic equation $\phi_s^j(\tau)$, and so in principle 
are time-dependent, since $\phi_s^j$ is. However, in practice we are 
interested in fluctuations about the stationary state, $\phi^*$, and so they 
lose their time dependence. They also only have a non-trivial spatial 
dependence through the presence of the discrete Laplacian, because the 
stationary state is homogeneous. Therefore the calculation can be considerably 
simplified by taking the spatial Fourier transform of Eqs.~(\ref{eq:langevin}) 
and (\ref{eq:lambda}). As discussed in Appendix \ref{MandB} this gives (see 
also~\cite{lug08})
\begin{equation}
\frac{d\xi^{\textbf{k}}_s}{d\tau}= \sum_{r} M_{sr}^{\textbf{k}}
\xi^{\textbf{k}}_{r} + \lambda^{\textbf{k}}_s(\tau),
\label{langevin_k}
\end{equation}
where
\begin{equation}
\langle \lambda_s^{\textbf{k}}(\tau)\lambda_r^{\textbf{k}'}(\tau')
\rangle = \mathcal{B}_{sr}^{\textbf{k}} \Omega a^d
\delta_{\textbf{k}+\textbf{k}',0} \delta(\tau - \tau'),
\label{lambda_k}
\end{equation}
and where $\textbf{k}$ is the wavevector. Here we have assumed that the 
micro-cells form a hypercubic lattice in $d-$dimensions with a lattice 
spacing $a$. The matrices $M^{\textbf{k}}$ and $\mathcal{B}^{\textbf{k}}$ are 
given by Eqs.~(\ref{FTofM})-(\ref{M^s_sr}) and 
Eqs.~(\ref{FTofB})-(\ref{B^s_sr}) respectively. However the important
point is that now $\textbf{k}$ is simply a label and the matrix structure
originating from the spatial nature of the problem has been lost. Thus both
$M^{\textbf{k}}$ and $\mathcal{B}^{\textbf{k}}$ are simply $k \times k$
matrices (recall that $k$ is the number of chemical species) and the analysis 
from now on is as in the non-spatial case~\cite{dau09}.

As we have already stressed in this paper, fluctuations about the stationary
state need to be taken into account, since they can be significant even if $N$
is quite large. The fact we can investigate these systematically is crucially
dependent on the linearity of Eq.~(\ref{langevin_k}) and that the
$M^{\textbf{k}}$ and $\mathcal{B}^{\textbf{k}}$ matrices are time-independent.
It means that we can straightforwardly take the temporal Fourier transform of
Eq.~(\ref{langevin_k}) to obtain
\begin{equation}
\sum^{k}_{r=1} \left( -i\omega \delta_{sr} - M_{sr}^{\textbf{k}} \right)
\tilde{\xi}^{\textbf{k}}_{r} (\omega) =
\tilde{\lambda}^{\textbf{k}}_{s}(\omega),
\label{FT_Lang}
\end{equation}
where $\tilde{f}$ denotes the temporal Fourier transform of the function
$f$. Defining the matrix $(-i\omega \delta_{sr} - M_{sr}^{\textbf{k}})$ to be
$\Phi^{\textbf{k}}_{sr}(\omega)$, the solution to Eq.~(\ref{FT_Lang}) is
\begin{equation}
\tilde{\xi}^{\textbf{k}}_{r} (\omega) = \sum^{k}_{s=1}
\left[ \Phi^{\textbf{k}}(\omega) \right]^{-1}_{rs}
\tilde{\lambda}^{\textbf{k}}_{s}(\omega).
\label{solution}
\end{equation}

From previous investigations, and the nature of the system, we expect that
the fluctuations about the stationary state (\ref{stationary}) will oscillate,
and will also be sustained and enhanced by a resonant
effect~\cite{mck05,dau09}. This is indeed what is seen. To investigate this
effect systematically we focus our attention on the power spectrum
$P_s(\textbf{k},\omega)$ of the fluctuations of species $s$,
\begin{eqnarray}
& & P_s(\textbf{k},\omega) \equiv
\left\langle | \xi^{\textbf{k}}_{s}(\omega) |^2 \right\rangle = \nonumber \\
& & \Omega a^d \sum^{k}_{r=1} \sum^{k}_{u=1}
\left[ \Phi^{\textbf{k}}(\omega) \right]^{-1}_{sr}
\mathcal{B}^{\textbf{k}}_{ru}
\left[ \Phi^{\textbf{k}\,\dag}(\omega) \right]^{-1}_{us}.
\label{power_spectrum}
\end{eqnarray}

The theoretical power spectrum can be found and plotted out, for any given 
choice of the chemical parameters, from Eq.~(\ref{power_spectrum}). To make 
contact with earlier investigations \cite{dau09}, and aiming at elucidating 
the spatial effects, we here solely focus on the choice $k=4$ and select 
$\eta=10$, $\beta=5/32$, and $\gamma=5/32$. When the $\alpha_s$ are set 
equal to zero, the communication between neighboring micro-cells is silenced, 
each spatial block behaving as an independent unit. Based on 
Eq.~(\ref{power_spectrum}), a temporal peak in the power spectrum is predicted 
to occur. The peak is approximately located at $\omega \simeq 4$, in agreement 
with the analysis developed in \cite{dau09}. Another simple limit is when
the $\alpha_s$ are made equal for all of the $k$ chemical species. In this 
case, the temporal peak gets progressively damped at large $\textbf{k}$, the 
effect being more pronounced the larger the values for the migration 
parameters. A similar phenomenon was also reported to occur in \cite{lug08}. 

\begin{figure*}[!htbp]
\begin{center}
\begin{tabular}{cc}
\includegraphics[width=80mm,height=50mm]{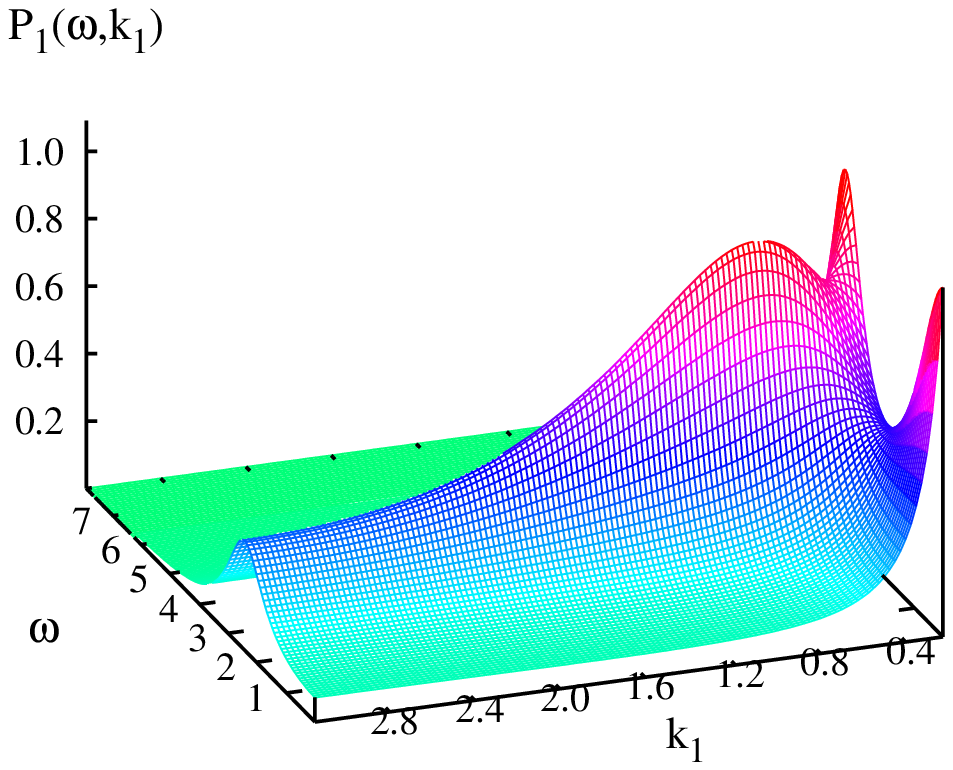} &
\includegraphics[width=80mm,height=50mm]{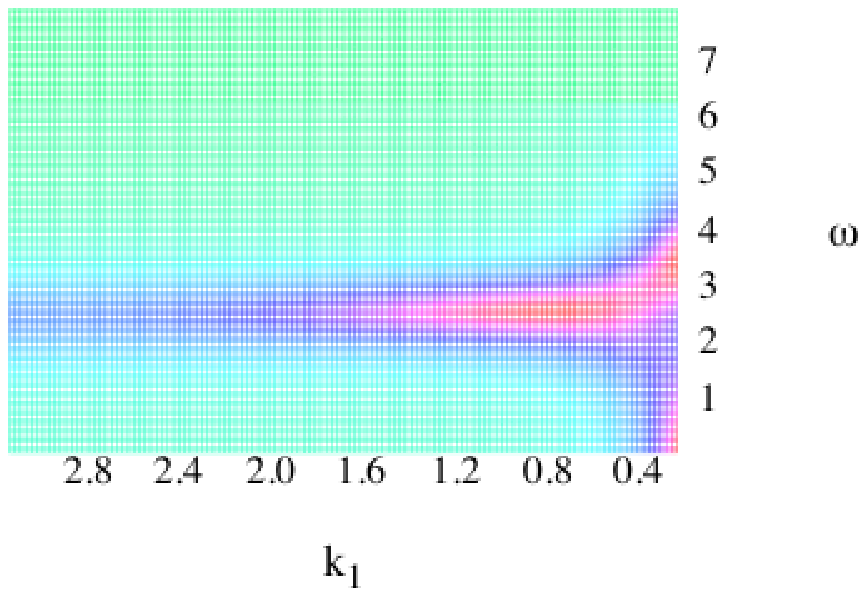}\\
\includegraphics[width=80mm,height=50mm]{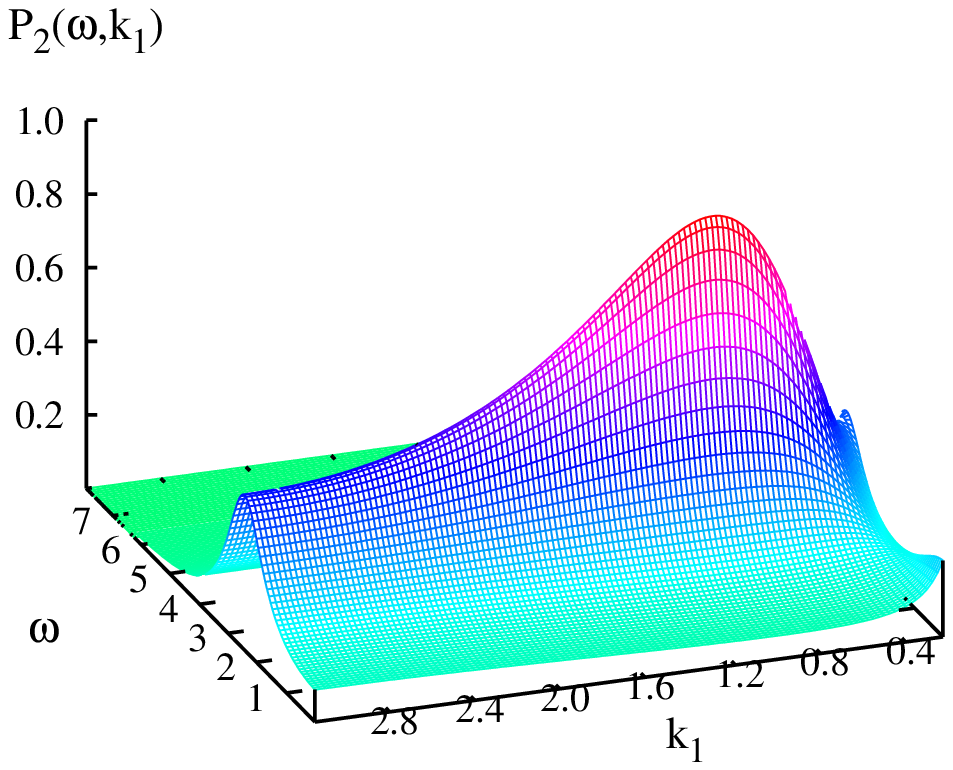} &
\includegraphics[width=80mm,height=50mm]{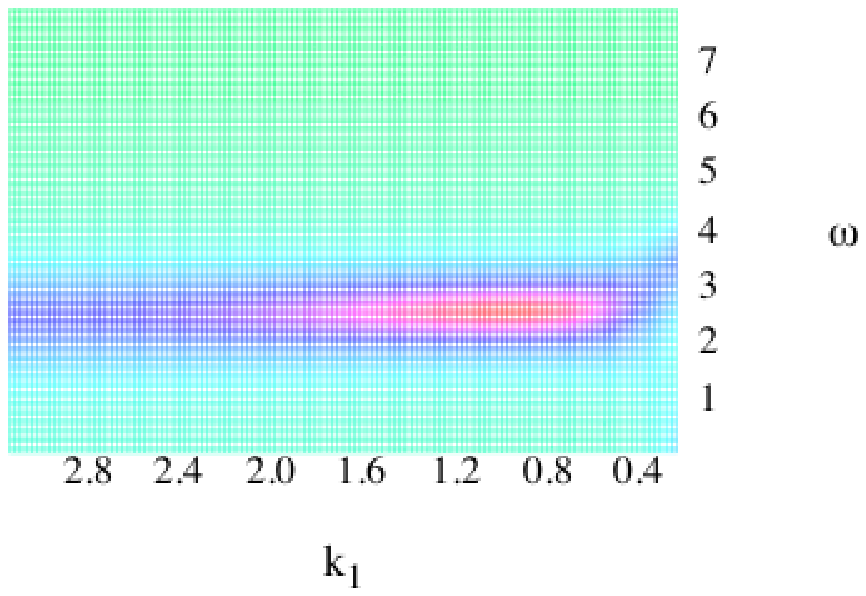}\\
\includegraphics[width=80mm,height=50mm]{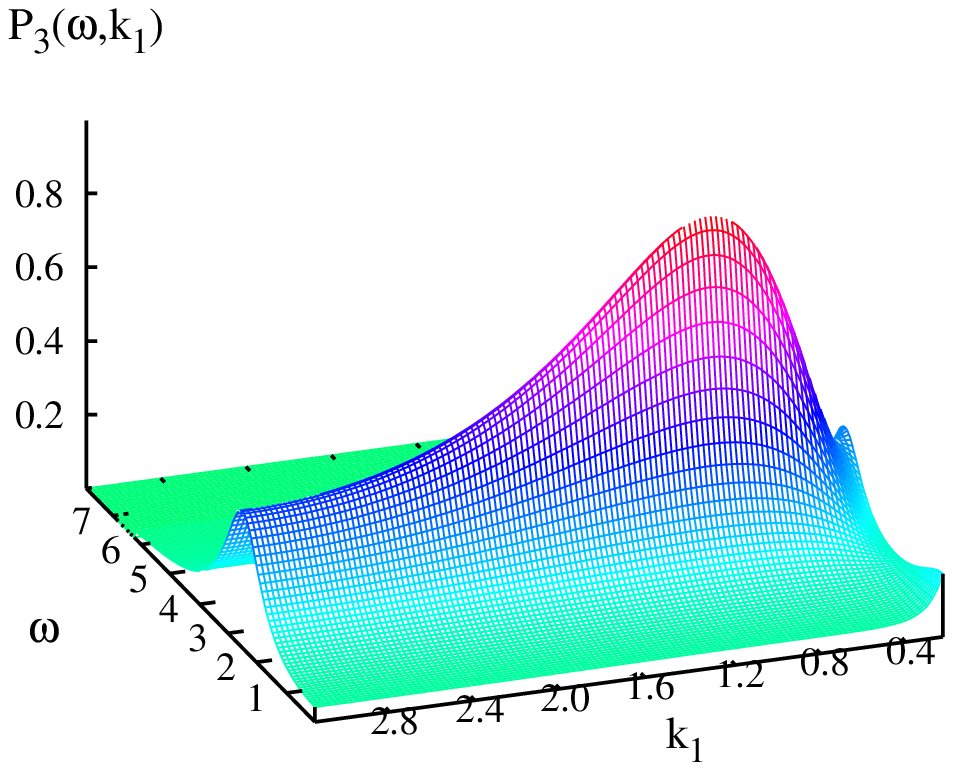} &
\includegraphics[width=80mm,height=50mm]{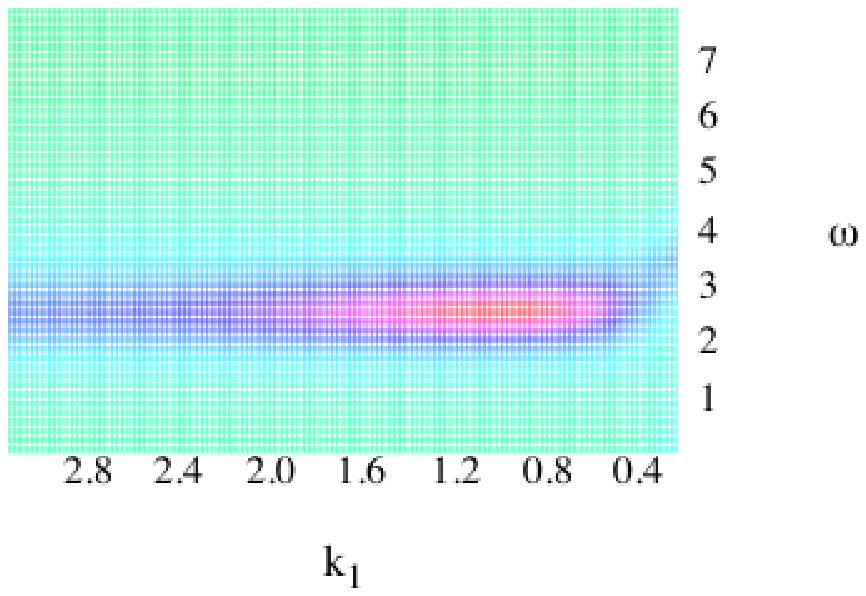}\\
\includegraphics[width=80mm,height=50mm]{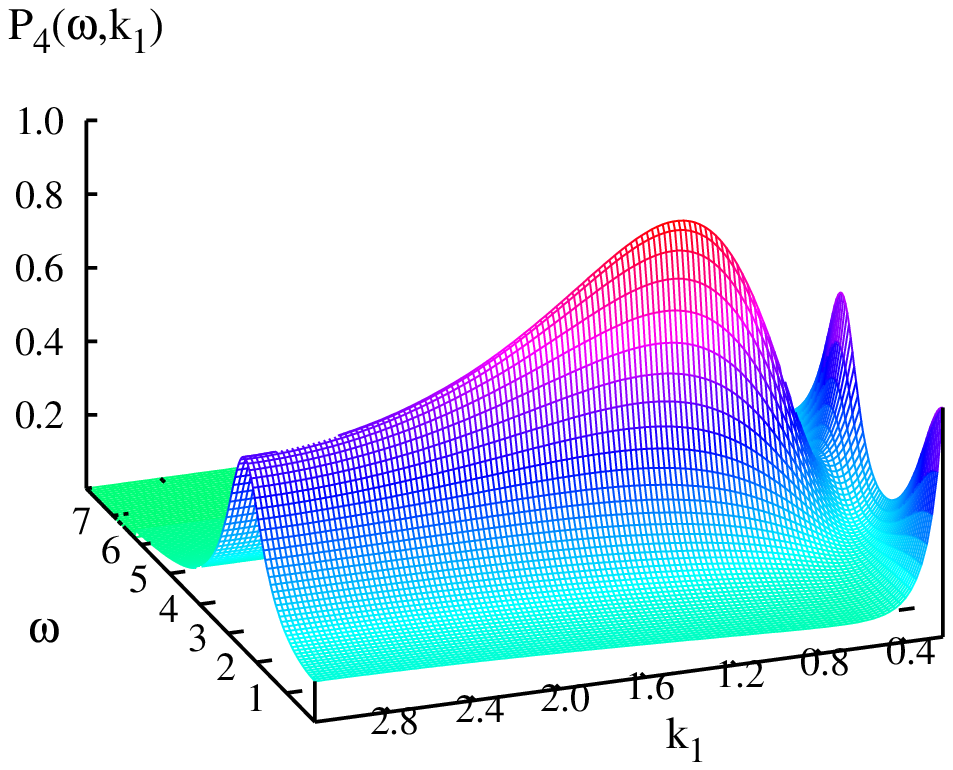} &
\includegraphics[width=80mm,height=50mm]{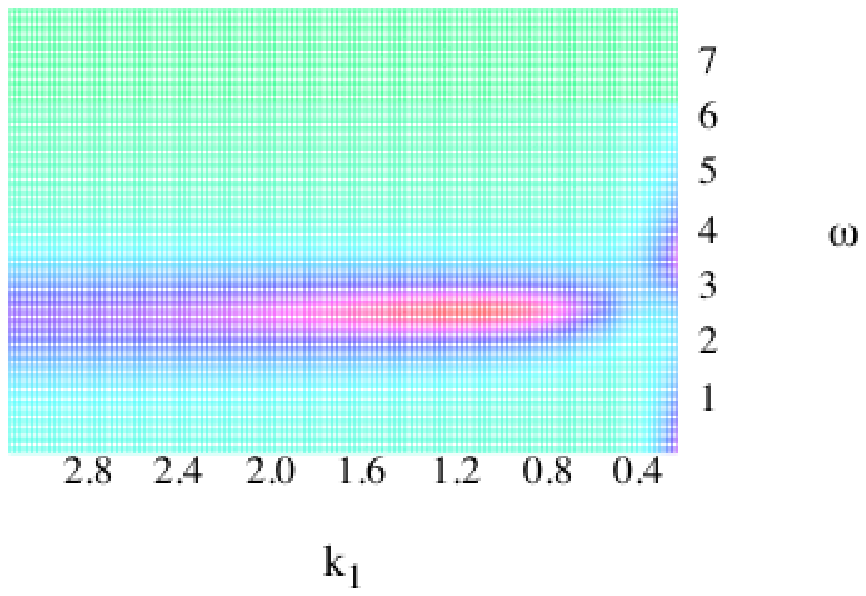}
\end{tabular}
\caption{(Color online) Power spectra calculated from 
Eq.~(\ref{power_spectrum}) for $k=4$ species and for a two-dimensional vesicle 
(one-dimensional periodic array of micro-cells). Here $N=5000$, $\Omega=256$,  
$\eta=10$, $\beta=5/32$, $\gamma=5/32$ and $\boldsymbol{\alpha} =
[100,0.001,1,500]$. Each pair of panels (the three-dimensional plot
and its two-dimensional projection) refers to a different chemical
species. A localized peak is displayed predicting the existence of
regular spatio-temporal patterns.} \label{fig:ana4}
\end{center}
\end{figure*}

More interestingly, in Fig.~\ref{fig:ana4}, we show the theoretical power 
spectrum Eq.~(\ref{power_spectrum}) for $\alpha_s$ that take different 
values for each of the species in the case of a two-dimensional vesicle 
(a one-dimensional periodic lattice of micro-cells, i.e.~$d=1$). The range 
of variation of the $\alpha_s$ covers several orders of magnitude, which in 
turn corresponds to assigning a significantly different degree of mobility 
to the species. Molecules characterized by large values of $\alpha_s$ will 
quickly diffuse, while those with smaller $\alpha_s$ are associated with 
relatively static, and presumably, more massive, species. A localized peak 
is clearly displayed suggesting that organized spatio-temporal patterns can 
spontaneously emerge, due to the inherent stochasticity of the system. From 
an inspection of Figs.~\ref{fig:ana4}, it is also evident that the power 
spectrum shows a clear peak for all four species. We found that making 
the $\alpha_s$ significantly different among species was a simple way to 
produce localized spatio-temporal patterns. We also found that they could be 
produced if (at least) one of the $\alpha_s$ was sufficiently large, when 
compared with the others.

\begin{figure*}[!htbp]
\begin{center}
\begin{tabular}{cc}
\includegraphics[width=80mm,height=50mm]{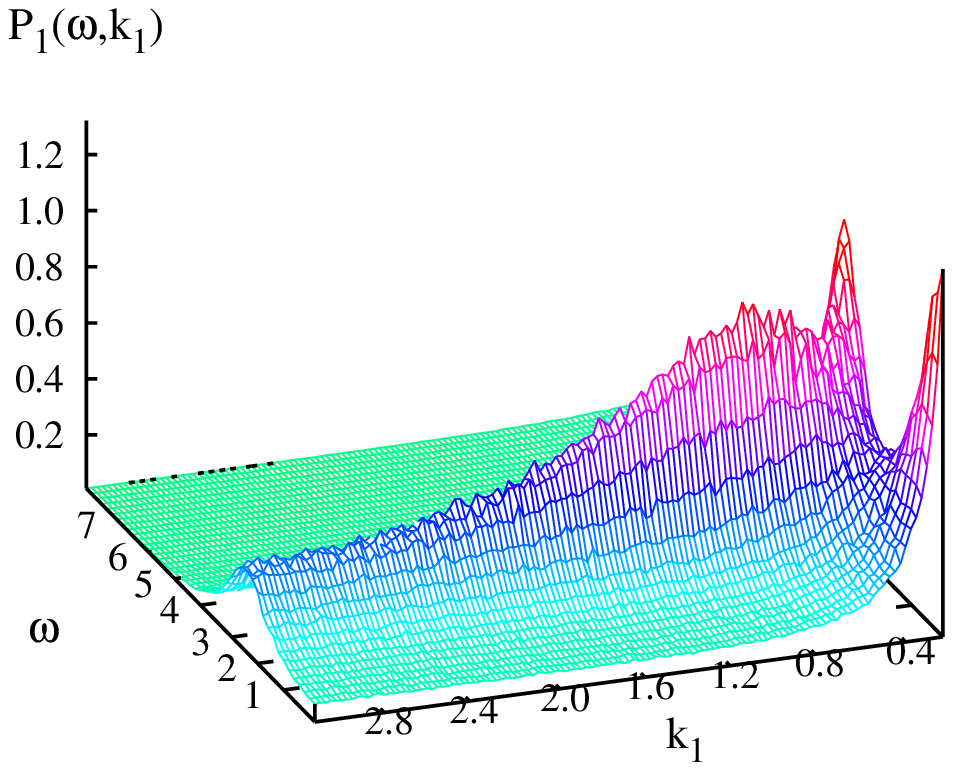} &
\includegraphics[width=80mm,height=50mm]{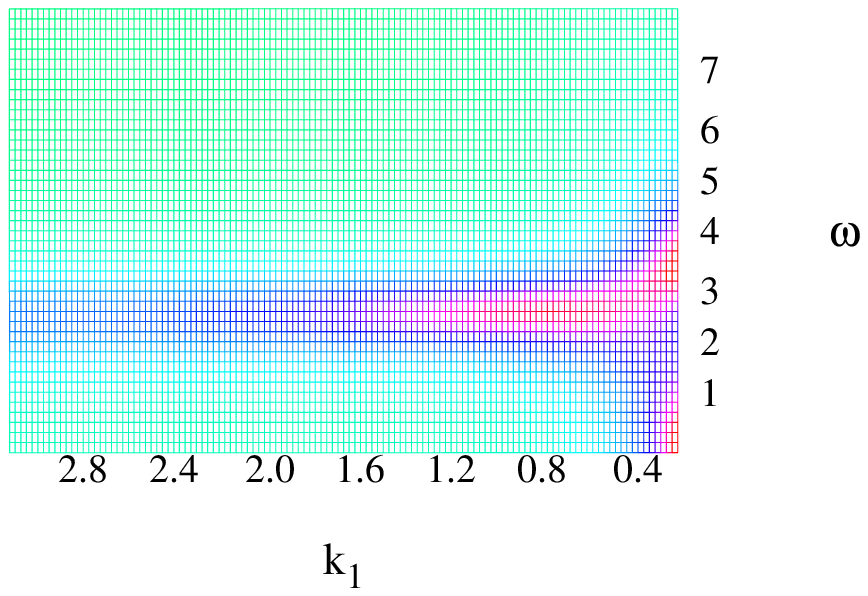}\\
\includegraphics[width=80mm,height=50mm]{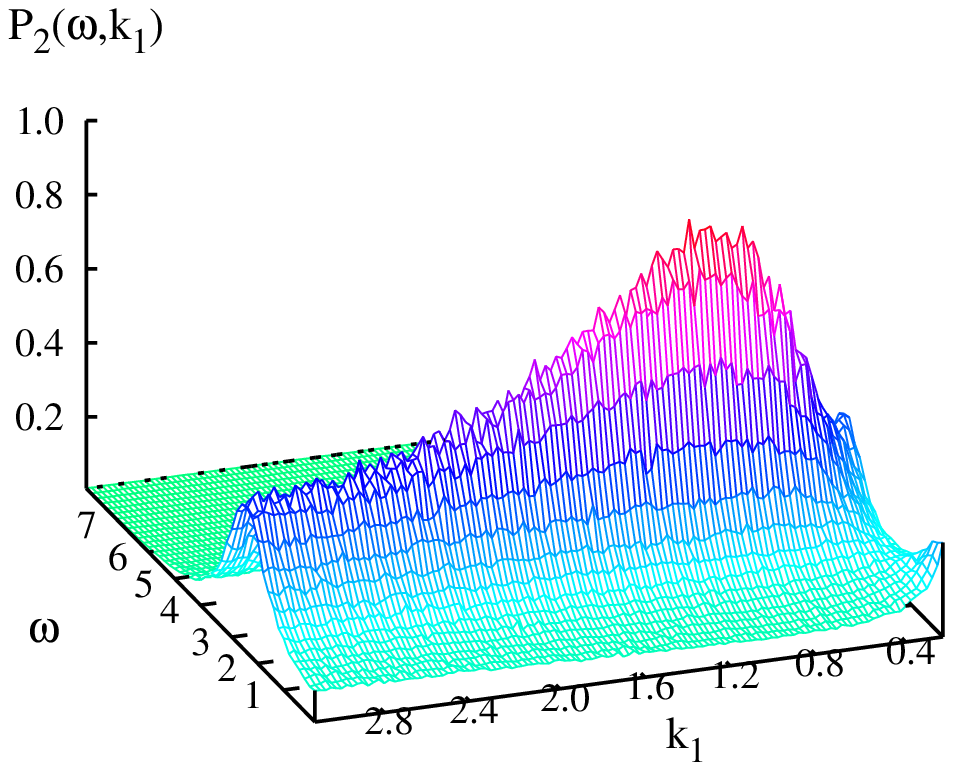} &
\includegraphics[width=80mm,height=50mm]{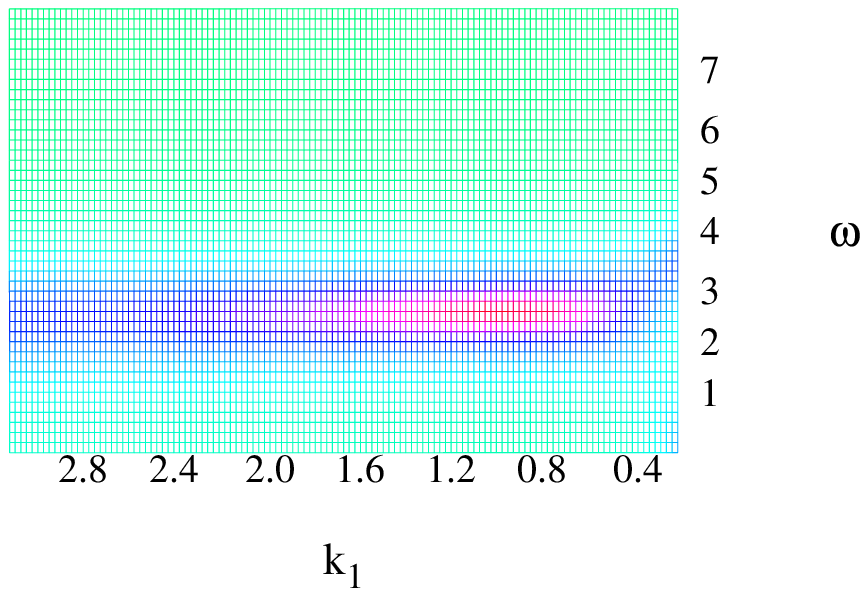}\\
\includegraphics[width=80mm,height=50mm]{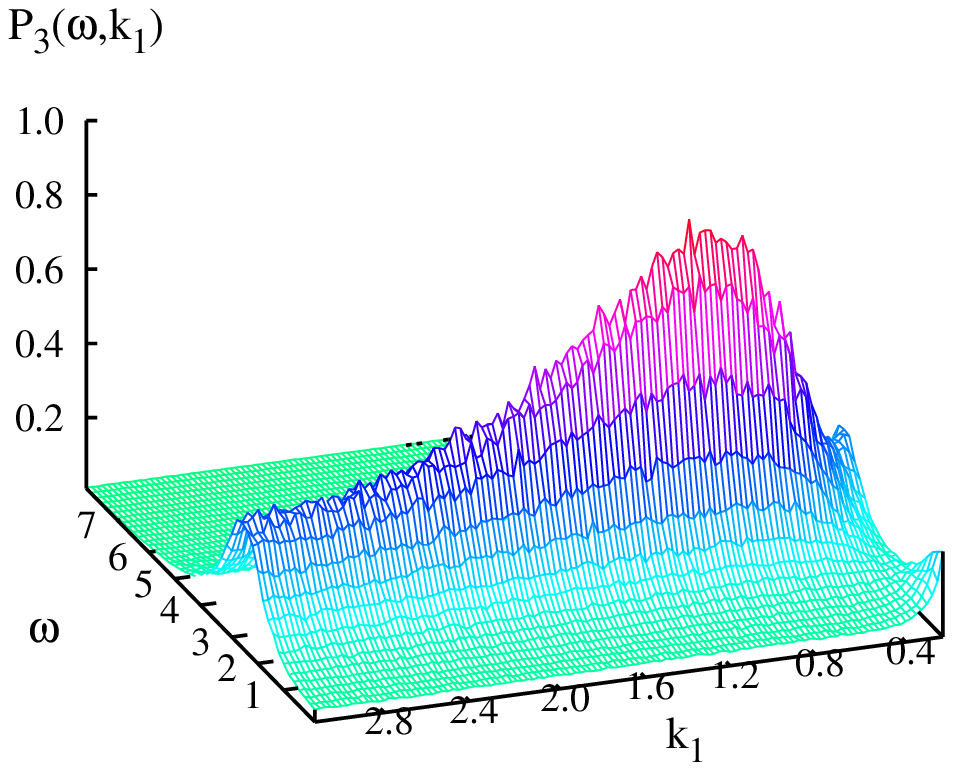} &
\includegraphics[width=80mm,height=50mm]{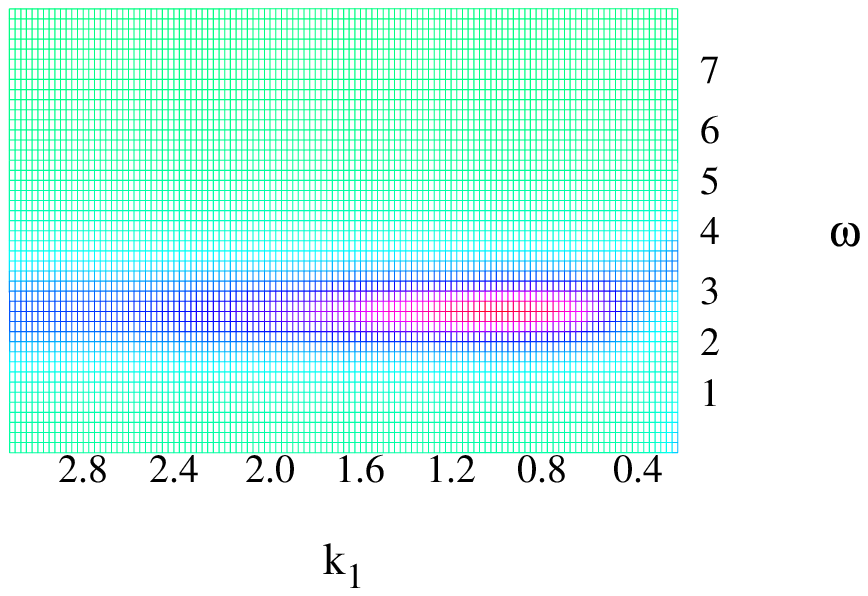}\\
\includegraphics[width=80mm,height=50mm]{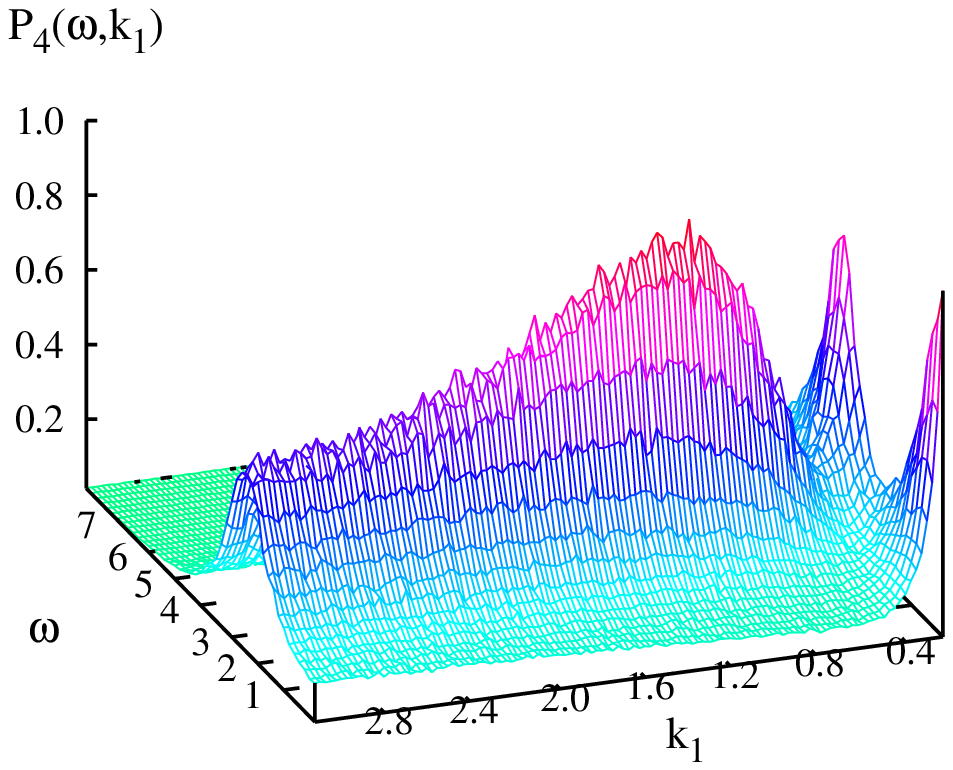} &
\includegraphics[width=80mm,height=50mm]{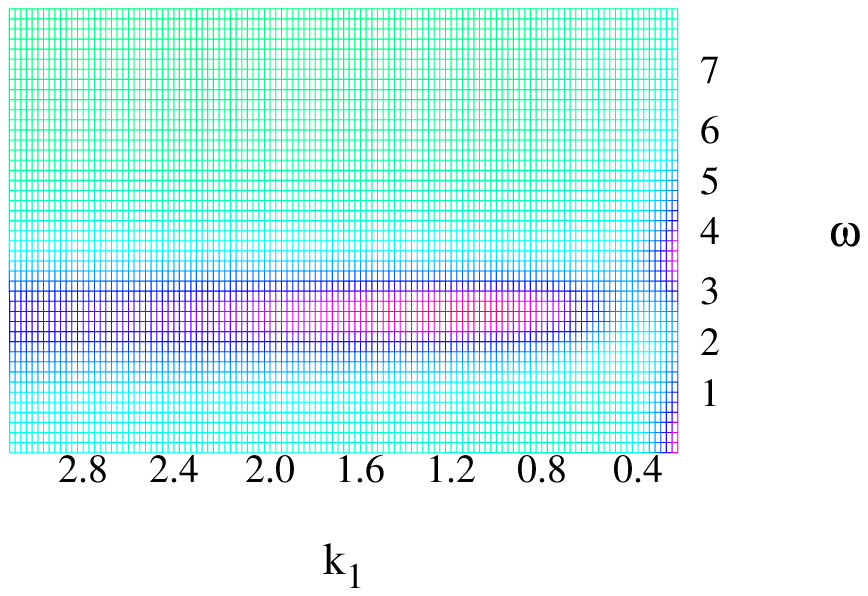}
\end{tabular}
\caption{(Color online) Numerically calculated power spectra obtained from
averaging 800 realizations. Stochastic simulations are performed via the 
Gillespie algorithm. Parameters are set as in Fig.~\ref{fig:ana4}.} 
\label{fig:num}
\end{center}
\end{figure*}

The conclusion of the above analysis, as well as the accuracy of the
approximations that have been employed, can be tested via direct numerical
simulations. By averaging over many realizations, we can calculate the power
spectra after Fourier transformation. Results of the simulation are displayed
in Fig.~\ref{fig:num} for the same choice of parameters as in 
Fig.~\ref{fig:ana4}. The correspondence between the profiles is excellent and
so confirms the correctness of our theoretical scheme.

In summary, we have unambiguously demonstrated that organized
spatio-temporal cycles can emerge in a simple model of protocells where the
constituents inside the vesicle interact via an autocatalytic scheme. As we
shall argue in the following, this finding provides a possible mechanism
to drive a dynamical synchronization between the duplication of genetic
material inside a protocell and the division of the vesicle membrane.

\section{Discussion}
\label{discussion}
In this paper we have investigated how the discreteness of the constituents in
an autocatalytic chemical reaction can lead to spatio-temporal oscillations.
The occurrence of temporal oscillations in such a setting, but without a
spatial element, has previously been studied~\cite{tog01,dau09}. Similarly,
such oscillations have been studied for a predator-prey system in a
spatial framework~\cite{lug08}, but the oscillations in this case did not
occur at a non-zero value of $\textbf{k}$. In this paper, we have combined and
generalized these treatments, and also put them into the context of vesicles,
which suggests an interesting consequence of the oscillatory behavior.

We can speculate that the natural tendency of the chemical constituents to
organize in regular spatio-temporal cycles can be instrumental in achieving
a degree of synchronization between the outer membrane of the vesicle and
the mixture of chemicals inside. In the context of protocells, these chemicals
undergoing autocatalytic reactions are to be interpreted as a primitive form
of genetic material. One would expect, as a minimal self-consistency
requirement, that within a stable population, a vesicle would split into
two when the chemical material contained within it had approximately doubled 
in size. It is tempting to postulate that such a property is a dynamical 
phenomenon, the density fluctuations acting as a positive feedback on the 
vesicle growth, so signaling when the constituents inside the vesicle are 
ready for the splitting to take place.

Now let us imagine that the vesicle containing the chemical species grows,
because of the inclusion of successive membrane constituents from the
environment in which it moves. Laboratory experiments indicate~\cite{lui06}
that a vesicle filled with water or solutes is kept in a turgid spherical
shape while growing by additional material of a similar kind flowing in from
the outside environment. It is believed that the vesicle remains spherical
until a thermodynamic instability sets in which distorts the
structure~\cite{fan08}, eventually leading to fissioning. Now suppose that
the vesicle is filled by a discrete population of chemical constituents,
which undergo an independent dynamics of the autocatalytic type. As
illustrated in this paper, the chemicals experience a first rapid evolution
towards the stationary state, where enhanced oscillations appear due to the
intrinsic finiteness of the interacting constituents. Such oscillations might
seed an instability~\cite{mac07a,mac07b}, which could resonate with the innate 
ability of the container to divide, so initiating the splitting process. These 
ideas could be extended to protocells, where enhanced oscillations could 
originate in the primitive genetic material. These oscillations could signal 
to the membrane that the genetic evolution had been virtually taken to 
completion and that the fission could now occur, so ensuring that the genetic 
material is passed on to the daughter protocells. This is a highly 
speculative suggestion, which calls for further investigation in the context 
of self-consistent formulations, where both the membrane and the genetic 
material are dynamically evolved. 

It is clear that the work presented here can be extended in various ways. The
nature of the lattice structure that is assumed can be generalized. For 
instance it is straightforward to include next-nearest neighbors, next-next 
nearest neighbor and so on. The analytical treatment is analogous, and the
results the same; only the form of the operator $\Delta_{\textbf{k}}$ changes.
Numerical simulations could also be performed in higher dimensions. In 
particular, a toroidal (donut-like) cell embedded in a three-dimensional 
space can be straightforwardly simulated. The inner volume of the cell is again
partitioned into micro-cells, and distinct diffusion rates are assigned to 
the radial and longitudinal directions. Preliminary simulations indicate that 
collective modes can develop giving rise to organized spatio-temporal 
dynamics \cite{dipa10}. 


\begin{acknowledgments}
 This work was partially funded by the HPC--EUROPA2 project (project number: 
228398) with the support of the European Commission --- Capacities 
Area --- Research Infrastructures.
\end{acknowledgments}

\appendix

\section{The van-Kampen expansion}
\label{VK}
In this Appendix we will give more details of the application of the van Kampen
system-size expansion to the master equation (\ref{master}). A general
discussion of the method is given in van Kampen's book~\cite{van07} and a
description of the application to a simple model showing sustained and
enhanced stochastic fluctuations is given in~\cite{mck05}. The calculations
given below build upon those carried out for the non-spatial version of the
model considered in this paper~\cite{dau09} and a spatial predator-prey
model~\cite{lug08}. We will occasionally refer back to these two papers below.

The starting point for the expansion in powers of $N^{-1/2}$ is the
ansatz (\ref{ansatz}). From this the following two results can be
derived~\cite{van07}. First, the left-hand side of the master
equation (\ref{master}) is given by
\begin{equation}
\frac{dP(\textbf{n},t)}{dt} =   \frac{\partial \Pi (\xi_m^i,t)}{\partial t} -
N^{\frac{1}{2}} \sum_{j=1}^{\Omega}
\sum_{s=1}^k \frac{\partial \Pi (\xi_m^i,t)}{\partial \xi_s^j}
\frac{d\phi_s^j}{dt},
\label{LHS}
\end{equation}
where $\Pi(\xi_m^i,t) \equiv P(n_m^i,t)$. Second, the step operator
${\cal E}^{\pm 1}_{s,j}$ may be expanded:
\begin{eqnarray}
{\cal E}^{\pm 1}_{s,j} &=& 1 \pm N^{-\frac{1}{2}}\frac{\partial}
{\partial \xi_s^j} + (2N)^{-1}\frac{\partial^2}{\partial (\xi_s^j)^2} +
\ldots \nonumber \\
&\equiv& 1 \pm N^{-\frac{1}{2}}\partial_{\xi_s^j} +
(2N)^{-1}\partial^2_{\xi_s^j} + \ldots.
\label{step_op_expand}
\end{eqnarray}

The right-hand side of the master equation may be also expanded. We begin
by defining new operators which are the coefficients of $N^{-1/2}$ and
$N^{-1}$ in the expansion of the particular combinations of the step operators
which appear in the model. These are
\begin{eqnarray*}
& & ({\cal E}_{s,j} {\cal E}_{s+1,j}^{-1} -1) \simeq
N^{-\frac{1}{2}}\Big( \partial_{\xi_s^j} - \partial_{\xi_{s+1}^j} \Big) \\
& & + \frac{1}{2}\Big[ N^{-\frac{1}{2}}\Big( \partial_{\xi_s^j}
- \partial_{\xi_{s+1}^j} \Big) \Big]^2
\equiv N^{-\frac{1}{2}} \widehat{L}_{1s} +
\frac{1}{2} N^{-1}\widehat{L}_{2s}, \\
& & ({\cal E}_{s,j} {\cal E}_{s,j'}^{-1} -1) \simeq
N^{-\frac{1}{2}}\Big( \partial_{\xi_s^j} - \partial_{\xi_{s}^{j'}} \Big) \\
& & + \frac{1}{2}\Big[ N^{-\frac{1}{2}}\Big( \partial_{\xi_s^j}
- \partial_{\xi_{s}^{j'}} \Big) \Big]^2
\equiv N^{-\frac{1}{2}} \widehat{L}_{1j} +
\frac{1}{2}N^{-1}\widehat{L}_{2j},
\end{eqnarray*}
where the operators $\widehat{L}_{1s}$ and $\widehat{L}_{2s}$ read
\begin{displaymath}
\widehat{L}_{1s} = \Big( \partial_{\xi_s^j} - \partial_{\xi_{s+1}^j} \Big),
\ \ \widehat{L}_{2s} = \Big( \partial_{\xi_s^j} -
\partial_{\xi_{s+1}^j}\Big)^{2},
\end{displaymath}
and where $\widehat{L}_{1j}$ and $\widehat{L}_{2j}$ read
\begin{displaymath}
\widehat{L}_{1j} = \Big( \partial_{\xi_s^j} - \partial_{\xi_{s}^{j'}} \Big),
\ \ \widehat{L}_{2j} = \Big( \partial_{\xi_s^j} - 
\partial_{\xi_{s}^{j'}}\Big)^2.
\end{displaymath}

In addition
\begin{displaymath}
({\cal E}_{s,j}^{-1} -1) \sim - N^{-\frac{1}{2}}\ \partial_{\xi_s^j} +
\frac{1}{2} N^{-1} \partial^2_{\xi_s^j},
\end{displaymath}
\begin{displaymath}
({\cal E}_{s,j} -1) \sim  N^{-\frac{1}{2}}\ \partial_{\xi_s^j} +
\frac{1}{2} N^{-1} \partial^2_{\xi_s^j}.
\end{displaymath}
For each of the three terms appearing in Eq.~(\ref{master}), namely the local,
migration and environmental terms, we can now identify the various
contributions in the van Kampen expansion: those of order $N^{-1/2}$, those
of order $N^{-1}$ involving a single derivative, and those of order $N^{-1}$
but involving two derivatives. We will examine these in turn.

\subsection{Right-hand side of the master equation: the $N^{-1/2}$ terms}

The contribution from ${\cal T}^{j}_{\rm loc}$, defined by Eq.~(\ref{local}),
is
\begin{displaymath}
\sum_s \frac{\eta_{s+1}}{\Omega} \widehat{L}_{1s}(\phi_{s}^j\phi_{s+1}^j).
\end{displaymath}
Using the definition of $\widehat{L}_{1s}$, shifting the sum on $s$ and
remembering that quantities with subscripts $k+1$ are to be identified with
those with subscripts $1$, we obtain
\begin{equation}
{\cal T}^{j (1)}_{\rm loc} = \frac{1}{\Omega} \sum_s (\eta_{s+1}
\phi_{s}^j\phi_{s+1}^j - \eta_{s} \phi_{s-1}^j\phi_{s}^j)\partial_{\xi_{s}^j},
\label{local_1}
\end{equation}
where the superscript $(1)$ indicates that this is only the contribution to
${\cal T}^{j}_{\rm loc}$ from terms of order $N^{-1/2}$. It should be noted
that Eq.~(\ref{local_1}) operates on $\Pi(\xi_m^i,t)$.

The contribution from ${\cal T}^{j j'}_{\rm mig}$, defined by
Eq.~(\ref{migration}) and using the definition of $\widehat{L}_{1j}$, is
\begin{displaymath}
\frac{2}{z\Omega} \sum_s \alpha_s \partial_{\xi_{s}^j}
\left( \phi_{s}^j(1-\sum_m\phi_{m}^{j'}) -
\phi_{s}^{j'}(1-\sum_m\phi_{m}^j)\right).
\end{displaymath}
To write this contribution in a way which naturally involves the Laplacian
operator we add to this sum two terms which add to zero, namely
\begin{displaymath}
0 = \phi_{s}^{j}\sum_m\phi_{m}^j - \phi_{s}^{j}\sum_m\phi_{m}^{j}.
\end{displaymath}
Summing the contribution over $j' \in j$ and introducing the discrete Laplacian
$\Delta f_s^j = (2/z) \sum_{j' \in j} (f_s^{j'} - f_s^j)$ we obtain
\begin{equation}
\sum_{j'} {\cal T}^{j j' (1)}_{\rm mig} = - \sum_s \frac{\alpha_s}{\Omega}
\Big( \Delta\phi_s^j(1-\sum_m \phi_m^j) + \phi_s^j \sum_m\Delta\phi_m^j\Big).
\label{migration_1}
\end{equation}

The contribution from ${\cal T}^{j}_{\rm env}$, defined by Eq.~(\ref{environ}),
is immediately found to be
\begin{equation}
{\cal T}^{j (1)}_{\rm env} = \sum_s \Big[\frac{\partial}{\partial \xi_s^j}
\Big(\frac{\gamma_{s}}{\Omega}\phi_s^j -
\frac{\beta_{s}}{\Omega}(1-\sum_m\phi_m^j)\Big) \Big].
\label{environ_1}
\end{equation}

Adding the three terms (\ref{local_1})-(\ref{environ_1}) together, and letting
them act on $\Pi(\xi_m^i,t)$ and summing over $j$, we may equate the resulting
expression to the order $N^{1/2}$ term in Eq.~(\ref{LHS}), after the
rescaling of time $\tau = t/(N\Omega)$. The resulting equation describes the
deterministic time-evolution of the species $s$ in micro-cell $j$ in the
limit $N \to \infty$, and is given by Eq.~(\ref{eq:first_order}) in the main
text.

\subsection{Right-hand side of the master equation: the $N^{-1}$ terms with
a single derivative}

These contributions are expressed in terms of the operators $\widehat{L}_{1s}$
and $\widehat{L}_{1j}$, and so are a function of the first derivatives in the
fluctuation variables. We proceed as we did for the terms of order $N^{-1/2}$.

The contribution from ${\cal T}^{j}_{\rm loc}$, defined by Eq.~(\ref{local}),
is
\begin{eqnarray}
& & {\cal T}^{j (2)}_{\rm loc} = \sum_{s=1}^k \frac{\eta_{s+1}}{\Omega}
\Big[\widehat{L}_{1s}\Big(\phi_s^j \xi_{s+1}^j
+ \phi_{s+1}^j\xi_s^j  \Big)    \Big] \nonumber \\
& = & \frac{1}{\Omega} \sum_{s=1}^k \Big[\frac{\partial}{\partial \xi_s^j}
\Big( \phi_s^j(\eta_{s+1} \xi_{s+1}^j - \eta_{s} \xi_{s-1}^j) \nonumber \\
& + & \xi_s^j(\eta_{s+1}\phi_{s+1}^j - \eta_{s}\phi_{s-1}^j)\Big) \Big],
\label{local_2}
\end{eqnarray}
where once again use has been made of the definition $\widehat{L}_{1s}$, the
cyclic nature of the species, and the sum on $s$ has been shifted. Here the
superscript $(2)$ indicates that this is only the contribution to
${\cal T}^{j}_{\rm loc}$ from terms of order $N^{-1}$ with a single derivative.

The contribution from ${\cal T}^{j j'}_{\rm mig}$, defined by
Eq.~(\ref{migration}), is
\begin{displaymath}
\frac{2}{z\Omega} \sum_s \alpha_s \Big[\widehat{L}_{1j}
\Big(\phi_s^j(-\sum_m\xi_m^{j'}) + \xi_s^j(1-\sum_m\phi_m^{j'}) \Big) \Big].
\end{displaymath}
Performing the same manipulations as before, but also inserting the identities
\begin{displaymath}
0 = \xi_{s}^{j}\sum_m\phi_{m}^j - \xi_{s}^{j}\sum_m\phi_{m}^{j}, \ \
0 = \phi_{s}^{j}\sum_m\xi_{m}^j - \phi_{s}^{j}\sum_m\xi_{m}^{j},
\end{displaymath}
gives, after summing over $j' \in j$,
\begin{eqnarray}
& & \sum_{j'} {\cal T}^{j j' (2)}_{\rm mig} = -\frac{1}{\Omega} \sum_s
\alpha_s \Big[\frac{\partial}{\partial \xi_s^j} \Big(\Delta\xi_s^j +
\xi_s^j \sum_m\Delta\phi_m^j \nonumber \\
& - & \Delta\phi_s^j\sum_m\xi_m^j + \phi_s^j\sum_m\Delta \xi_m^j
- \Delta \xi_s^j\sum_m\phi_m^j \Big) \Big].
\label{migration_2}
\end{eqnarray}

Finally, the contribution from ${\cal T}^{j}_{\rm env}$, defined by
Eq.~(\ref{environ}), is immediately found to be
\begin{equation}
{\cal T}^{j (2)}_{\rm env} = \sum_s \Big[\frac{\partial}{\partial \xi_s^j}
\Big(\frac{\gamma_{s}}{\Omega}\xi_s^j + \frac{\beta_{s}}{\Omega}
\sum_m\xi_m^j \Big) \Big].
\label{environ_2}
\end{equation}

\subsection{Right-hand side of the master equation: the $N^{-1}$ terms with
two derivatives}

These terms are expressed in terms of the operator $\widehat{L}_{2s}$ and
$\widehat{L}_{2j}$, and so are a function of the second order derivatives in
the fluctuation variables. We proceed as we did in the two previous cases.

The contribution from ${\cal T}^{j}_{\rm loc}$ is
\begin{eqnarray}
& & {\cal T}^{j (3)}_{\rm loc} = \frac{1}{\Omega} \sum_s \eta_{s+1}
\frac{1}{2}\widehat{L}_{2s}(\phi_s^j \phi_{s+1}^j) \nonumber \\
& = & \frac{1}{ 2\Omega} \sum_s \eta_{s+1} (\phi_s^j \phi_{s+1}^j)
\Big(\frac{\partial^2}{\partial (\xi_s^j)^2} \nonumber \\
& & + \frac{\partial^2}{\partial (\xi_{s+1}^j)^2} -
2\frac{\partial^2}{\partial \xi_s^j \partial \xi_{s+1}^j} \Big).
\label{local_3}
\end{eqnarray}

The contribution from ${\cal T}^{j j'}_{\rm mig}$ is
\begin{eqnarray}
& & {\cal T}^{j j' (3)}_{\rm mig} = \frac{1}{z\Omega} \sum_s
\alpha_s \frac{1}{2}\Big[\widehat{L}_{2j}(\phi_s^j (1-\sum_m\phi_m^{j'}))
\Big] = \nonumber \\
& & \alpha_s  \frac{1}{z\Omega} \sum_s \Big(\phi_s^j
(1-\sum_m\phi_m^{j'}) \Big) \Big[ \frac{\partial} {\partial \xi_s^j}
- \frac{\partial}{\partial \xi_s^{j'}}\Big]^{2}. \label{migration_3}
\end{eqnarray}

Finally, the contribution from ${\cal T}^{j}_{\rm env}$ is found to be
\begin{equation}
{\cal T}^{j (3)}_{\rm env} = \frac{1}{2} \sum_s \Big(\frac{\beta_{s}}{\Omega}
(1-\sum_m\phi_m^j) + \frac{\gamma_{s}}{\Omega}\phi_s^j\Big)\frac{\partial^2}
{\partial(\xi_s^j)^2}.
\label{environ_3}
\end{equation}

Adding the six terms (\ref{local_2})-(\ref{environ_3}) together, and letting
them act on $\Pi(\xi_m^i,t)$ and summing over $j$, we may equate the resulting
expression to the order one term in Eq.~(\ref{LHS}), after the rescaling of
time $\tau = t/(N\Omega)$. The resulting equation gives the stochastic
time-evolution of the species $s$ in micro-cell $j$. It takes the form of a
Fokker-Planck equation which we now examine.

\section{The form of the matrices $M$ and $\mathcal{B}$}
\label{MandB}
To write down the differential equation for $\Pi(\xi_s^j,\tau)$, it is
convenient to combine the indices $s$ and $j$, labeling the species and
micro-cells respectively, by a single index $p$. To do this we imagine the
$k\Omega$-dimensional vector with components $\xi_s^j$ as an ordered sequence
of $\Omega$ vectors, each of $k$ components. This can be achieved by
defining $p = (j-1)k + s$, so that the component $\xi_p$ represents the
fluctuations associated with the $s$-th species in the $j$-th micro-cell.

Now the terms in Eqs.~(\ref{local_2})-(\ref{environ_2}) take the form of
a single derivative of $\xi_p$ acting on a linear combination of $\xi_l$,
$l=1,\ldots,k\Omega$ and the terms in Eqs.~(\ref{local_3})-(\ref{environ_3})
are a linear combination of second order derivatives. Therefore using this
more compact notation the Fokker-Planck takes the form
\begin{eqnarray}
\frac{\partial \Pi}{\partial \tau} &=& -\sum_p
\frac{\partial}{\partial \xi_p} \Big[\mathcal{A}_p(\boldsymbol{\xi})
\Pi\Big] + \frac{1}{2} \sum_{l,p} \mathcal{B}_{lp}
\frac{\partial^2 \Pi}{\partial \xi_l \partial \xi_p}, \nonumber \\
\label{eq:FP_sp}
\end{eqnarray}
where the matrix $\mathcal{A}$ can be re-written as
\begin{equation}
\mathcal{A}_p(\boldsymbol{\xi}) = \sum_l M_{pl} \xi_{l}.
\label{eq:A}
\end{equation}

So to specify the Fokker-Planck equation we need to give the form of
the two $(k\Omega)\times(k\Omega)$ matrices $M$ and $\mathcal{B}$.
We first note that although they do not depend on the fluctuations
$\xi_{p}(\tau)$, they do depend on the solution of the deterministic
differential equation (\ref{eq:first_order}), as well as on the
reaction rates $\eta_{s}, \alpha_{s}, \beta_{s}$ and $\gamma_s$.
However since we are only interested in the fluctuations about the
stationary state $\phi^*$ given by Eq,~(\ref{stationary}), and since
we obtained this solution under the assumption that $\eta_{s},
\beta_{s}$ and $\gamma_s$ were independent of $s$, we take the two
matrices to only depend on $\phi^{*}, \eta, \beta, \gamma$ and
$\alpha_s$. An inspection of Eqs.~(\ref{local_2})-(\ref{environ_3})
reveals that the only spatial dependence in $M$ and $\mathcal{B}$
originates from the discrete Laplacian. This suggests that if we
introduce spatial Fourier transforms, we should be able to
diagonalize $M$ and $\mathcal{B}$, and so be left with matrices only
in the species space. This is most easily carried out by not
continuing to work with the Fokker-Planck equation (\ref{eq:FP_sp}),
but instead with the equivalent Langevin equation~\cite{gar04,ris89}
\begin{equation}
\frac{d\xi_p}{d\tau}= \mathcal{A}_p (\boldsymbol{\xi}) + \lambda_p(\tau),
\label{langevin}
\end{equation}
where
\begin{equation}
\langle \lambda_p(\tau)\lambda_q(\tau')
= \mathcal{B}_{pq}\delta(\tau - \tau'),
\label{lambda}
\end{equation}
and where the noise term, $\lambda_p(\tau)$, in Eq.~(\ref{langevin}) is 
Gaussian with zero mean. This is Eq.~(\ref{eq:langevin}) in the main text, 
but using the single index notation.

We follow the conventions and methods of~\cite{lug08} for the spatial Fourier
transforms. For simplicity, we shall assume that the lattice is a
$d-$dimensional hypercubic lattice, with lattice spacing $a$. Then the
Fourier transform, $f_s^{\bf k}$, of a function $f_s^{\bf j}$, is defined by
\begin{equation}
f_s^{\bf k} = a^{d} \sum_{\bf j} e^{ - i {\bf k}.a{\bf j}}\,f_s^{\bf j},
\label{FTdef}
\end{equation}
where we have now written the lattice site label ${\bf j}$ as a vector to
emphasize the $d-$dimensional nature of the transform. We may now take
the spatial Fourier transform of the matrix $M$. Since the only spatial
dependence is through the discrete Laplacian, we may decompose
Eq.~(\ref{eq:A}) as follows:
\begin{equation}
A^{\textbf{j}}_s = \sum_{\textbf{j}'} \sum_{r} M^{\textbf{jj}'}_{sr}
\xi^{\textbf{j}'}_{r}
= \sum_{r} \left[ M^{(NS)}_{sr} \xi^{\textbf{j}}_{r} + M^{(SP)}_{sr}
\Delta \xi^{\textbf{j}}_{r} \right],
\label{decompose}
\end{equation}
where the two $k \times k$ matrices $M^{(NS)}$ and $M^{(SP)}$ will be specified
below. It is now straightforward to take the spatial Fourier transform of
Eq.~(\ref{decompose}) to obtain~\cite{lug08}
\begin{equation}
A^{\textbf{k}}_s = \sum_{r} \left[ M^{(NS)}_{sr} + M^{(SP)}_{sr}
\Delta_{\textbf{k}} \right] \xi^{{\textbf{k}}}_{r},
\label{FTofA}
\end{equation}
where $\Delta_{\textbf{k}}$ is the Fourier transform of the discrete Laplacian
and is given by
\begin{equation}
\Delta_{\textbf{k}}= \frac{2}{d} \sum_{\gamma =1}^d
\left[ \cos(\textrm{k}_{\gamma}a) - 1 \right].
\label{Delta_k}
\end{equation}
Care should be taken not to confuse the components of the wavevector
$\textbf{k}$, and $k$, the number of chemical species. We have denoted the
$\gamma$-th component of the wave vector $\textbf{k}$ by $\textrm{k}_{\gamma}$
to help avoid this confusion.

The quantity within the square brackets in Eq.~(\ref{FTofA}) is the
spatial Fourier transform of the matrix $M$. It is diagonal in
$\textbf{k}-$space and so depends on the single label $\textbf{k}$.
We may therefore write it as
\begin{equation}
M^{\textbf{k}}_{sr} = M^{(NS)}_{sr} + M^{(SP)}_{sr} \Delta_{\textbf{k}},
\label{FTofM}
\end{equation}
where the two matrices $M^{(NS)}$ and $M^{(SP)}$ may be read off from
Eqs.~(\ref{local_2})-(\ref{environ_2}), and are given by
\begin{eqnarray}
\label{M^n_ss}
&& M^{(NS)}_{ss} = - \beta - \gamma  \\
&& M^{(NS)}_{sr} =\left\{ \begin{array}{ll}
-\eta \phi^* - \beta, & \qquad \textrm{if } r=s+1 \\
\eta \phi^* - \beta, & \qquad \textrm{if } r=s-1 \\
- \beta, & \qquad \textrm{if } |s-r|>1,
\end{array} \right.
\label{M^n_sr}
\end{eqnarray}
and
\begin{eqnarray}
\label{M^s_ss}
&& M^{(SP)}_{ss} = \alpha_{s} \left[ 1 + (1-k)\phi^* \right] \\
&& M^{(SP)}_{sr} = \alpha_{s} \phi^{*} \ \ \textrm{if}\ s \neq r.
\label{M^s_sr}
\end{eqnarray}
The matrix $M^{(NP)}$ is exactly the one found in the non-spatial version of 
the model~\cite{dau09}, which is why we have attached the label `$NS$' to it to
signify the non-spatial contribution to $M$. The spatial, or `$SP$',
contribution is simply $M^{(SP)}\Delta_{\textbf{k}}$.

To take the Fourier transform of the matrix $\mathcal{B}$, we note that out
of the three terms --- given by Eqs.~(\ref{local_3})-(\ref{environ_3}) --- from
which this matrix is constructed, the only non-trivial spatial dependence
comes from Eq.~(\ref{migration_3}). We display the contribution containing
this dependence by noting the following relation:
\begin{eqnarray}
& & \sum_{\textbf{j}} \sum_{\textbf{j}' \in \textbf{j}}
\Big[ \frac{\partial}{\partial \xi_s^{\textbf{j}}} -
\frac{\partial}{\partial \xi_s^{\textbf{j}'}}\Big]^2 = \nonumber \\
& & 2\sum_{\textbf{j}} \sum_{\textbf{j}'}
\Big[z\frac{\partial^2}{\partial(\xi_s^\textbf{j})^2}
\delta_{\textbf{jj}'} - \frac{\partial^2}
{\partial\xi_s^{\textbf{j}} \partial \xi_s^{\textbf{j}'}}
J_{<\textbf{jj}'>} \Big],
\label{lapl}
\end{eqnarray}
where $J_{<\textbf{jj}'>}$ is equal to $1$ if $\textbf{j}'$ and $\textbf{j}$
are nearest neighbors, and zero otherwise. The part of the $\mathcal{B}$
matrix corresponding to the expression (\ref{lapl}) is
$(2z \delta_{\textbf{jj}'} - 2J_{<\textbf{jj}'>})$ which has Fourier transform
\begin{displaymath}
a^d \left[ 2z - 4 \sum_{\gamma =1}^d \cos(\textrm{k}_{\gamma}a) \right]
= -z a^d \Delta_{\textbf{k}},
\end{displaymath}
using Eq.~(\ref{Delta_k}) and $z=2d$. Therefore we may express the matrix
$\mathcal{B}$ in Fourier space in a similar way to Eq.~(\ref{FTofM}):
\begin{equation}
\mathcal{B}^{\textbf{k}}_{sr} = \mathcal{B}^{(NS)}_{sr} +
\mathcal{B}^{(SP)}_{sr} \Delta_{\textbf{k}},
\label{FTofB}
\end{equation}
where the two $k \times k$ matrices $\mathcal{B}^{(NS)}$ and 
$\mathcal{B}^{(SP)}$ may be read off from 
Eqs.~(\ref{local_3})-(\ref{environ_3}), and are given by
\begin{eqnarray}
\label{B^n_ss}
&& \mathcal{B}^{(NS)}_{ss} = a^{d} \left[ \beta (1 - k\phi^* ) + \gamma \phi^*
+ 2\eta \left( \phi^* \right)^{2} \right] \\
&& \mathcal{B}^{(NS)}_{sr} =\left\{ \begin{array}{ll}
- a^d \eta \left( \phi^* \right)^{2} , & \qquad \textrm{if } r=s+1 \\
- a^d \eta \left( \phi^* \right)^{2} , & \qquad \textrm{if } r=s-1 \\
0 & \qquad \textrm{if } |s-r|>1,
\end{array} \right.
\label{B^n_sr}
\end{eqnarray}
and
\begin{eqnarray}
\label{B^s_ss} && \mathcal{B}^{(SP)}_{ss} = - 2 a^d \alpha_{s} \phi^*
\left( 1-k\phi^* \right) \\
&& \mathcal{B}^{(SP)}_{sr} = 0 \ \ \textrm{if}\ s \neq r.
\label{B^s_sr}
\end{eqnarray}
Once again, the matrix $\mathcal{B}^{(NS)}$ is exactly the one found in the
non-spatial version of the model~\cite{dau09}, up to a factor of $a^d$, which 
is why we have attached the label `$NS$' to it. The spatial contribution is
$\mathcal{B}^{(SP)}\Delta_{\textbf{k}}$.

\end{document}